Dynamic Modeling and Simulation of a Rotational Inverted Pendulum

# Dynamic Modeling and Simulation of a Rotational Inverted Pendulum

**J L Duart, B Montero, P A Ospina and E González.**

E-mail: juan.duarte03@ustabuca.edu.co, brayan.montero@ustabuca.edu.co, paolo.ospina@ustabuca.edu.co, edwin.gonzalez@ustabuca.edu.co

**Abstract**. This paper presents an alternative way to the dynamic modeling of a rotational inverted pendulum using the classic mechanics known as Euler-Lagrange allows to find motion equations that describe our model. It also has a design of the basic model of the system in SolidWorks software, which based on the material and dimensions of the model provides some physical variables necessary for modeling. In order to verify the theoretical results, It was made a contrast between the solutions obtained by simulation SimMechanics-Matlab and the system of equations Euler-Lagrange, solved through ODE23tb method included in Matlab bookstores for solving equations systems of the type and order obtained. This article comprises a pendulum trajectory analysis by a phase space diagram that allows the identification of stable and unstable regions of the system.

## 1. Introduction

Nowadays, the underactuated mechanical systems [1] are generating interest among researchers of modern control theories. This interest is that these systems have similar problems to those found in industrial applications, such as external shocks and/or non-linear behavior in some conditions of operation. The rotational inverted pendulum [2] is a clear example of a mechanical underactuated system; this is a mechanism of two degrees of freedom (DOF) and two rotational joints. It consists of three main elements: a motor, a rotational arm and a pendulum [3]. The motor shaft is connected to one end of the rotational arm making this fully rotate on a horizontal plane; the other end of the arm has connected the pendulum that freely rotates 360 degrees in a vertical plane. Despite being a purely academic level, this system is helpful to study, apply and analyze different modeling strategies.

Some industrial applications [4] that present the previously mentioned drawbacks and behavior, are in fields such as robotics, robots balance, biped robots [5], robotic arms; aerospace, positioning rocket; telecommunications, satellite positioning; transportation, Segway [6,7], stability of ships and submarines, iBot, Self-balancing unicycle; construction, bridge cranes; and field monitoring, drones.

To carry out this work, initially a literature review of similar projects [8,9,10] was conducted in order to determine the appropriate dimensions for the prototype design in SolidWorks CAD software. Then it was performed continuously dynamic modeling of the system and using the Lagrangian mechanics as a method, two equations of motion were obtained, due to each one of the two links comprising, the inverted pendulum has a degree of freedom (DOF). To validate the equations obtained, multiple simulations were made in the Matlab software in order to observe their behavior. Thus it was possible to verify in a graphic mode, the waveform of the position and angular velocity of the arm and the pendulum.

A fundamental part in the simulation stage, was to bring the CAD model made in SolidWorks to SimMechanics [11] Matlab environment. At the same time, the equations obtained using the Lagrangian mechanics were resolved with the ODE23tb method (Ordinary Differential Equations) belonging to the Matlab software. The use of block diagrams in Simulink allowed a representation of the system. This was intended in order to develop an overlap of results of both simulations and establish similarities and/or differences in the waveform of the variables of interest.

Finally, the graph of the effective potential of the system allowed tracing paths and obtain the phase space diagram [12] in which they can study the critical points that correspond to the maximum and minimum of the potential.

In section II the design of the basic model is presented with their respective physical variables. In section III It provides a brief definition of systems underactuated. Section IV presents the advantage of making use of Euler-Lagrange formalism. In Section V it shows the dynamic modeling of proposed rotational inverted pendulum. In section VI activities are defined for perform proposed simulations as a method of validating results. In section VII the results obtained are shown with a respective analysis of them. Finally, section VIII contains the relevant conclusions of the article.

## 2. Prototype Proposed

This section presents the basic design created in SolidWorks CAD software with the following features:

**Table 1.** Variable System

| Physical characteristic | Symbology |
|---|---|
| Pendulum mass | $m_1$ |
| Arm length | $L_0$ |
| Pendulum length | $L_1$ |
| Location of the center of mass of the pendulum | $l_1$ |
| Moment of inertia of the arm | $I_0$ |
| Moment of inertia of the pendulum | $I_1$ |
| angular position of the arm | $\theta_0$ |
| angular position of the pendulum | $\theta_1$ |

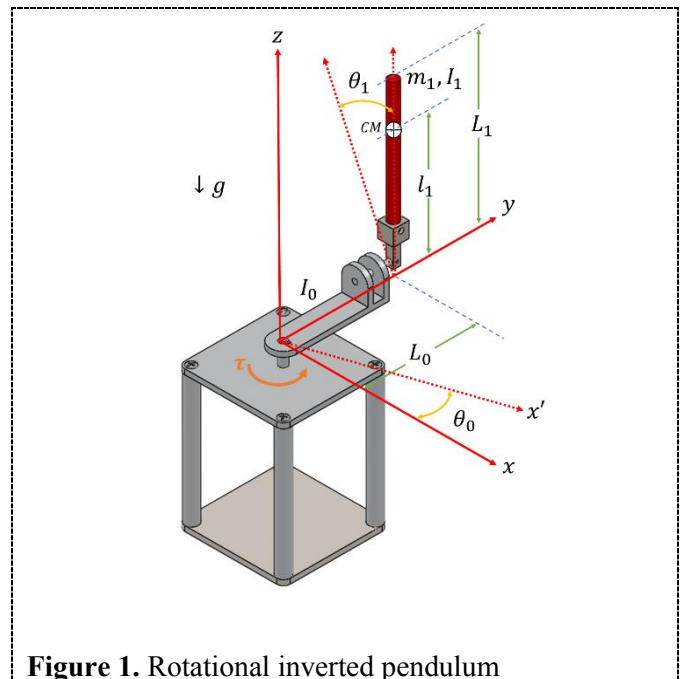

**Figure 1.** Rotational inverted pendulum

## 3. Underactuated Systems

Direct and indirect actions are two concepts that are acquired in the study of mechanisms. The first consists of movement of elements by action of an actuator, while the second consists of the action of motion transmitted by another interconnected element. Such movements are known as degrees of freedom, so that mechanical systems or mechanisms can be classified depending on the number of degrees of freedom and the number of actuators. The fully actuated mechanical systems are those having the same number of degrees of freedom and actuators. Underactuated mechanical systems are those with fewer actuators than degrees of freedom [13]. It is important to highlight the advantages of underactuated systems, since if they do not have advantages over fully actuated mechanical systems, it will not make sense its development. The main advantages present in underactuated systems are: energy saving and control efforts. However, these systems are intended to perform the same functions of fully actuated systems without their disadvantages.

## 4. Euler-Lagrange

The dynamic equations of any mechanical system can be obtained from the known classical mechanics (Newton), the drawback of this formalism is the use of the variables in vector form, complicating considerably the analysis when increasing the joints or there are rotations present in the system. In these cases, it is favorable to employ the Lagrange equations, which have formalism of scale, facilitating the analysis for any mechanical system.

In order to use Lagrange equations, it is necessary to follow four steps:

*1)* Calculation of kinetic energy.
*2)* Calculation of the potential energy.
*3)* Calculation of the Lagrangian.
*4)* Solve the equations.

Where the kinetic energy can be both rotationally and translational, this form of energy may be a function of both the position and the speed $K(q(t), \dot{q}(t))$.

The potential energy is due to conservative forces as the forces exerted by springs and gravity, this energy is in terms of the position $U(q(t))$.

The Lagrangian is defined as:

$$L = K - U \tag{1}$$

So the Lagrangian in general terms is defined as follows:

$$L\big(q(t), \dot{q}(t)\big) = K\big(q(t), \dot{q}(t)\big) - U(q(t)) \tag{2}$$

Finally the Euler-Lagrange equations for a $n$ degrees of freedom system of is defined as follows:

$$\frac{d}{dt}\left(\frac{\partial L(q,\dot{q})}{\partial \dot{q}_i}\right) - \frac{\partial L(q,\dot{q})}{\partial q_i} = \tau_i \tag{3}$$

Where $i = 1, \ldots, n$, $\tau_i$ are forces or externally exercised pairs (actuators) at each joint, besides nonconservative forces such as friction, resistance to movement of an object within a fluid and generally those that depend on time or speed. It will be obtained an equal number of dynamic equations and degrees of freedom.

## 5. Modeling of Rotational Inverted Pendulum

Lagrangian system modeling was performed as shown in Figure 1. It is necessary to make an energy analysis. Therefore, initially the kinetic energy of each link is analyzed, so it can be identified which kinetic energies (rotational and translational) were present in each link.

### 5.1. Kinetic Energy

The kinetic energy of the system consist of a translational and rotational component for the pendulum and rotational component for the arm,

$$K = \frac{1}{2}mv^2 + \frac{1}{2}I\omega^2 \tag{4}$$

Where m is the mass of the body, $v$ the linear velocity, $I$ the moment of inertia, $\omega$ the angular velocity and $K$ the kinetic energy. In this case there are two bodies, the arm and the pendulum. The kinetic energy of the arm is:

$$K_0 = \frac{1}{2}I_0\dot{\theta}_0{}^2 \tag{5}$$

The kinetic energy of the pendulum is:

$$K_1 = \frac{1}{2}m_1 v_1{}^2 + \frac{1}{2}I_1\dot{\theta}_1{}^2 \tag{6}$$

The total energy of the system is:

$$K_T = \frac{1}{2}I_0\dot{\theta}_0{}^2 + \frac{1}{2}m_1 v_1{}^2 + \frac{1}{2}I_1\dot{\theta}_1{}^2 \tag{7}$$

### 5.2. Potential Energy

This system only store gravitational potential energy in the pendulum.

$$U = mgh \tag{8}$$

The arm has in its nature a rotational movement in a horizontal plane, therefore it does not have height change in its center of mass, providing equal 0 component in equation (8) resulting in a potential energy zero .

The potential energy of the pendulum is,

$$U_1 = m_1 g l_1 (\cos\theta_1 - 1) \tag{9}$$

Where $g$ represents the value of gravity. The total potential energy of the system is (9),

$$U_T = m_1 g l_1 (\cos\theta_1 - 1) \tag{10}$$

### 5.3. Position of the pendulum
Because the pendulum is a rigid body, the required position is the its center of mass,

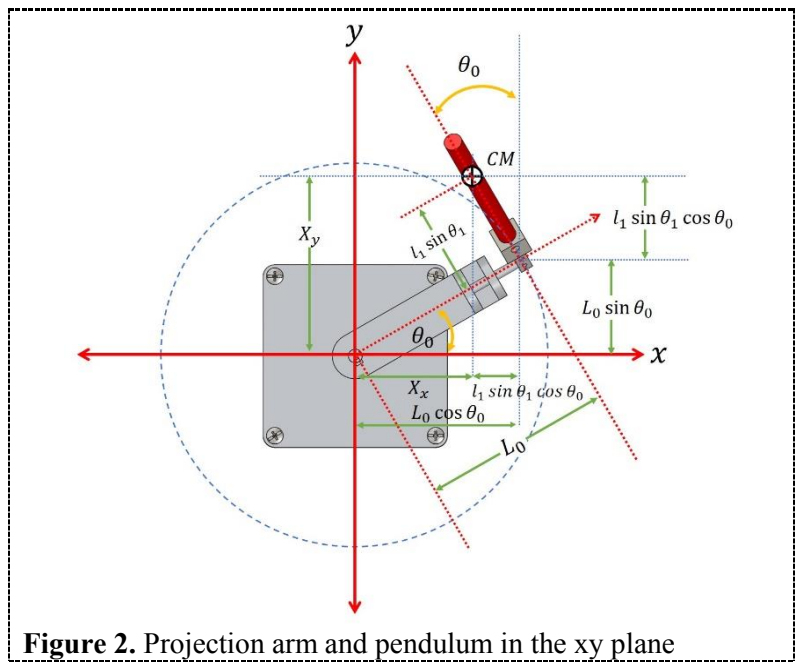

**Figure 2.** Projection arm and pendulum in the xy plane

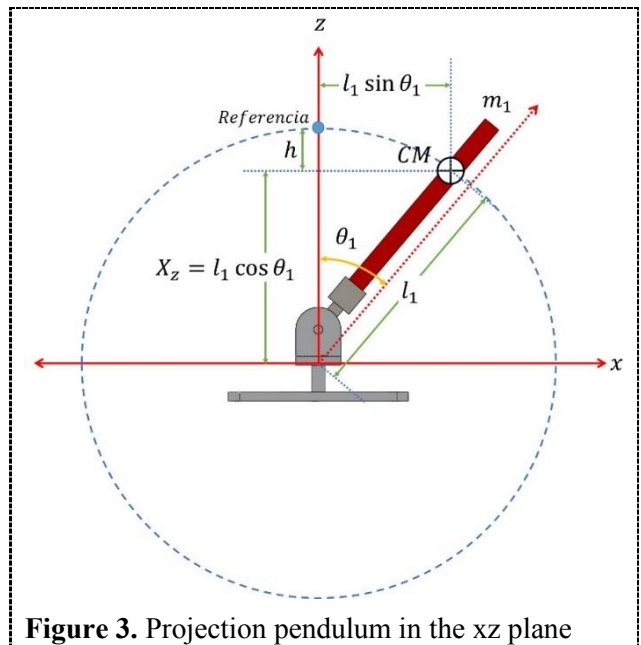

**Figure 3.** Projection pendulum in the xz plane

Then, it shows components the position of the center of mass.

$$X_x = L_0 \cos\theta_0 - l_1 \sin\theta_1 \sin\theta_0, \qquad (11)$$

$$X_y = L_0 \sin\theta_0 + l_1 \sin\theta_1 \cos\theta_0, \qquad (12)$$

$$X_z = l_1 \cos\theta_1, \qquad (13)$$

The position of the center of mass,

$$X_{CM} = [X_x \quad X_y \quad X_z]^T \qquad (14)$$

### 5.4. Linear Speed of the center of mass

The speed is defined as the derivative of the position with respect to time.

$$\frac{d}{dt}(X_{CM}) = V_{CM} \qquad (15)$$

$$\dot{X}_{CM} = V_{CM} = [\dot{X}_x \quad \dot{X}_y \quad \dot{X}_z]^T \qquad (16)$$

The velocity components are obtained by differentiating each component position (11), (12) and (13) respectively,

$$\dot{X}_x = -\dot{\theta}_0 L_0 \sin\theta_0 - l_1(\dot{\theta}_0 \sin\theta_1 \cos\theta_0 + \dot{\theta}_1 \cos\theta_1 \sin\theta_0) \qquad (17)$$

$$\dot{X}_y = \dot{\theta}_0 L_0 \cos\theta_0 + l_1(\dot{\theta}_1 \cos\theta_1 \cos\theta_0 - \dot{\theta}_0 \sin\theta_1 \sin\theta_0) \qquad (18)$$

$$\dot{X}_z = -\dot{\theta}_1 l_1 \sin\theta_1 \qquad (19)$$

The linear velocity of the center of mass can be expressed as follows vector form,

$$V_{CM}^2 = \begin{bmatrix} \dot{X}_x & \dot{X}_y & \dot{X}_z \end{bmatrix} \begin{bmatrix} \dot{X}_x \\ \dot{X}_y \\ \dot{X}_z \end{bmatrix} = \dot{X}_x{}^2 + \dot{X}_y{}^2 + \dot{X}_z{}^2 \tag{20}$$

They are calculated separately each of the components to the square of the velocity of the center of mass.

$$\dot{X}_x{}^2 = \dot{\theta}_0{}^2 L_0{}^2 \sin^2\theta_0 - 2\big(\dot{\theta}_0 L_0 \sin\theta_0\big)\big(l_1\big(\dot{\theta}_0 \sin\theta_1 \cos\theta_0 + \dot{\theta}_1 \cos\theta_1 \sin\theta_0\big)\big) + l_1{}^2\big(\dot{\theta}_0 \sin\theta_1 \cos\theta_0 + \dot{\theta}_1 \cos\theta_1 \sin\theta_0\big)^2 \tag{21}$$

$$\dot{X}_y{}^2 = \dot{\theta}_0{}^2 L_0{}^2 \cos^2\theta_0 + 2\big(\dot{\theta}_0 L_0 \cos\theta_0\big)\big(l_1\big(\dot{\theta}_1 \cos\theta_1 \cos\theta_0 - \dot{\theta}_0 \sin\theta_1 \sin\theta_0\big)\big) + l_1{}^2\big(\dot{\theta}_1 \cos\theta_1 \cos\theta_0 - \dot{\theta}_0 \sin\theta_1 \sin\theta_0\big)^2 \tag{22}$$

$$\dot{X}_z{}^2 = \dot{\theta}_1{}^2 l_1{}^2 \sin^2\theta_1 \tag{23}$$

Finally the following expression is obtained for the linear velocity.

$$V_{CM}^2 = L_0{}^2 \dot{\theta}_0{}^2 + l_1{}^2\big(\dot{\theta}_1{}^2 + \dot{\theta}_0{}^2 \sin^2\theta_1\big) + 2L_0 l_1 m_1 \dot{\theta}_0 \dot{\theta}_1 \cos\theta_1 \tag{24}$$

### 5.5. Energy System
The total energy of the system is,

$$K_T = K_0 + K_1 \tag{25}$$

Substituting (5) and (6) in (25) is obtained,

$$K_T = \frac{1}{2} I_0 \dot{\theta}_0{}^2 + \frac{1}{2} I_1 \dot{\theta}_1{}^2 + \frac{1}{2} m_1 V_{CM}^2 \tag{26}$$

Total kinetic energy is obtained by substituting equation (24) in (26)

$$K_T = \frac{1}{2} I_0 \dot{\theta}_0{}^2 + \frac{1}{2} I_1 \dot{\theta}_1{}^2 + \frac{1}{2} m_1 \{L_0{}^2 \dot{\theta}_0{}^2 + l_1{}^2\big(\dot{\theta}_1{}^2 + \dot{\theta}_0{}^2 \sin^2\theta_1\big) + 2L_0 l_1 m_1 \dot{\theta}_0 \dot{\theta}_1 \cos\theta_1\} \tag{27}$$

The potential energy of the system is shown in equation (10)

### 5.6. Euler-Lagrange Equations
The Lagrangian of the system is,

$$L = K_T - U_T \tag{28}$$

Substituting (10) and (27) (28) the Lagrangian of the system is obtained,

$$\begin{aligned} = \frac{1}{2} I_0 \dot{\theta}_0{}^2 + \frac{1}{2} I_1 \dot{\theta}_1{}^2 + \frac{1}{2}\big(L_0{}^2 m_1 \dot{\theta}_0{}^2\big) + \frac{1}{2}\big(l_1{}^2 m_1 \dot{\theta}_1{}^2\big) + \frac{1}{2}\big(l_1{}^2 m_1 \dot{\theta}_0{}^2 \sin^2\theta_1\big) \\ + L_0 l_1 m_1 \dot{\theta}_0 \dot{\theta}_1 \cos\theta_1 + m_1 g l_1 (1 - \cos\theta_1) \end{aligned} \tag{29}$$

As there are two degrees of freedom (DOF) the Euler-Lagrange equations have the following form,

$$\frac{d}{dt}\left(\frac{\partial L}{\partial \dot{\theta}_0}\right) - \frac{\partial L}{\partial \theta_0} = \tau \tag{30}$$

$$\frac{d}{dt}\left(\frac{\partial L}{\partial \dot{\theta}_1}\right) - \frac{\partial L}{\partial \theta_1} = 0 \tag{31}$$

Where $\tau$ is the torque of the motor. Solving (30) and (31) we obtain the equations of motion which are given by,

$$I_0\ddot{\theta}_0 + L_0{}^2 m_1\ddot{\theta}_0 + \left\{l_1{}^2 m_1\left(\ddot{\theta}_0\sin^2\theta_1 + 2\dot{\theta}_0\dot{\theta}_1\sin\theta_1\cos\theta_1\right)\right\} + \left\{L_0 l_1 m_1\left(\ddot{\theta}_1\cos\theta_1 - \dot{\theta}_1{}^2\sin\theta_1\right)\right\} = \tau \tag{32}$$

$$I_1\ddot{\theta}_1 + l_1{}^2 m_1\ddot{\theta}_1 + L_0 l_1 m_1\ddot{\theta}_0\cos\theta_1 - l_1{}^2 m_1\dot{\theta}_0{}^2\sin\theta_1\cos\theta_1 - m_1 g l_1\sin\theta_1 = 0 \tag{33}$$

Where (32) is the equation of motion of the arm (33) of the pendulum.

## 6. Simulation

Although the Euler-Lagrange formalism ensures a high degree of approximation of the mathematical models, is essential do comparisons to validate these results. For verification of modeling the following steps are followed:

*1)* Represent the system equations in the state space.
*2)* Define an experiment with initial conditions, natural interactions and external forces.
*3)* Import the CAD model of SolidWorks in SimMechanics-Matlab.
*4)* Add the necessary blocks to obtain the desired graphic model and applying external forces.
*5)* Simulating experiment.
*6)* Export the results of SimMechanics to Workspace Matlab.
*7)* Implement a block diagram Simulink-Matlab to solve the equations.
*8)* Export the solutions to the equations to Workspace Matlab.
*9)* Graphing and overlay solutions.

As can be seen in the steps above, the simulation of the model was divided into two stages: first, simulate the CAD model initially designed and the second, implement the equations obtained.

## 7. Results

Then, it presents each of the steps mentioned in the previous section.

*1)* Starting from [14],

$$M(q)\ddot{q} + C(q,\dot{q})\dot{q} + G(q) = \tau \tag{34}$$

So the equation (34) is the dynamic equation for mechanical systems of n degrees of freedom (DOF). Where $M$ is the matrix of inertia of the system, $C$ is the centrifugal and Coriolis matrix, $G$ the vector of gravity and $\tau$ external forces.

Taking the equations of motion (32) and (33) and replacing in (34) is obtained representation in state space system,

$$\begin{bmatrix} I_o + m_1 L_0{}^2 + l_1{}^2 m_1\sin^2\theta_1 & L_0 l_1 m_1\cos\theta_1 \\ L_0 l_1 m_1\cos\theta_1 & I_1 + m_1 l_1{}^2 \end{bmatrix}\begin{bmatrix}\ddot{\theta}_0 \\ \ddot{\theta}_1\end{bmatrix}$$
$$+ \begin{bmatrix} 2l_1{}^2 m_1\sin\theta_1\cos\theta_1\dot{\theta}_1 & -L_0 l_1 m_1\sin\theta_1\dot{\theta}_1 \\ -l_1{}^2 m_1\sin\theta_1\cos\theta_1\dot{\theta}_0 & 0 \end{bmatrix}\begin{bmatrix}\dot{\theta}_0 \\ \dot{\theta}_1\end{bmatrix} + \begin{bmatrix} 0 \\ -g l_1 m_1\sin\theta_1 \end{bmatrix} = \begin{bmatrix}\tau \\ 0\end{bmatrix} \tag{35}$$

The matrix $M(q)$ is important for the dynamic modeling and for the design of controllers. This matrix has a great relationship with the kinetic energy, also the inertia matrix is a symmetric, positive and square matrix of $n \times n$, whose elements depend only on the generalized coordinates.

Centrifugal and Coriolis matrix $C(q, \dot{q})$ it is important in the study of stability in control systems, mechanical systems, among others. This matrix is square of $n \times n$ and has dependence in its elements of the generalized coordinates and velocities.

The gravity vector $G(q)$ it is present in mechanical systems without counterweights or springs, in turn is in systems with displacement off the horizontal plane. This vector is of $n \times 1$ and has only reliance on joint positions.

*2)* It defined that the system would have the initial conditions shown in Figure 5, in addition to being subject to effects of gravity and to a torque step 0.2 seconds in the end of the arm that connects to the motor shaft. Finally, a simulation interval 5 seconds was established.

*3)* Was imported the CAD model in SimMechanics with the following code line:

*mech_import('CADModel_Pendulum.xml')*

The initial position of the system is Figure 5 a).

*4)* It was necessary add a few block the diagram SimMechanics, since the CAD model is only under the effect of gravity and not have some kind of movement, it blocks provide the step of torque arm included to start rotating, besides adding blocks to the sensing of angular displacement and velocities in an interval of 5 second of test.

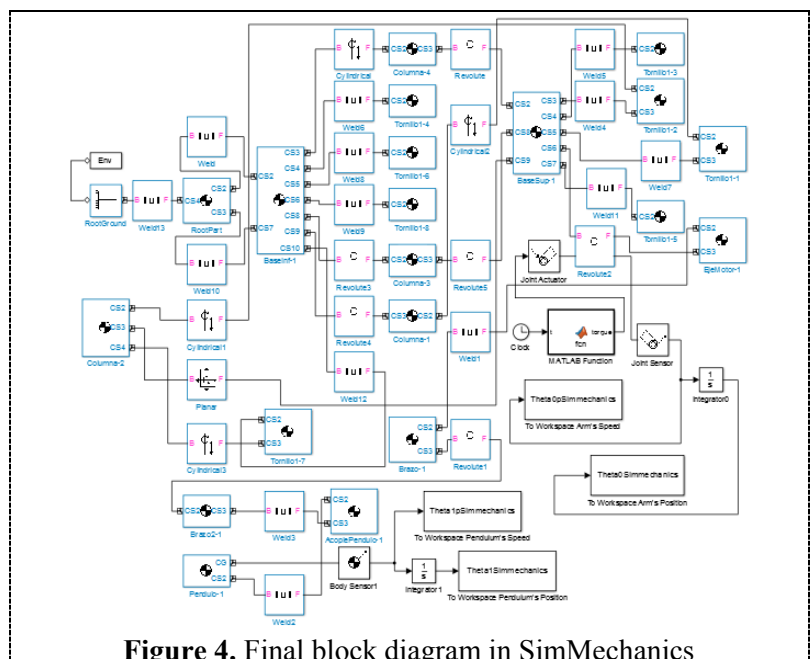

**Figure 4.** Final block diagram in SimMechanics

*5)* Some moments of the position of the system were taken in the time interval defined.

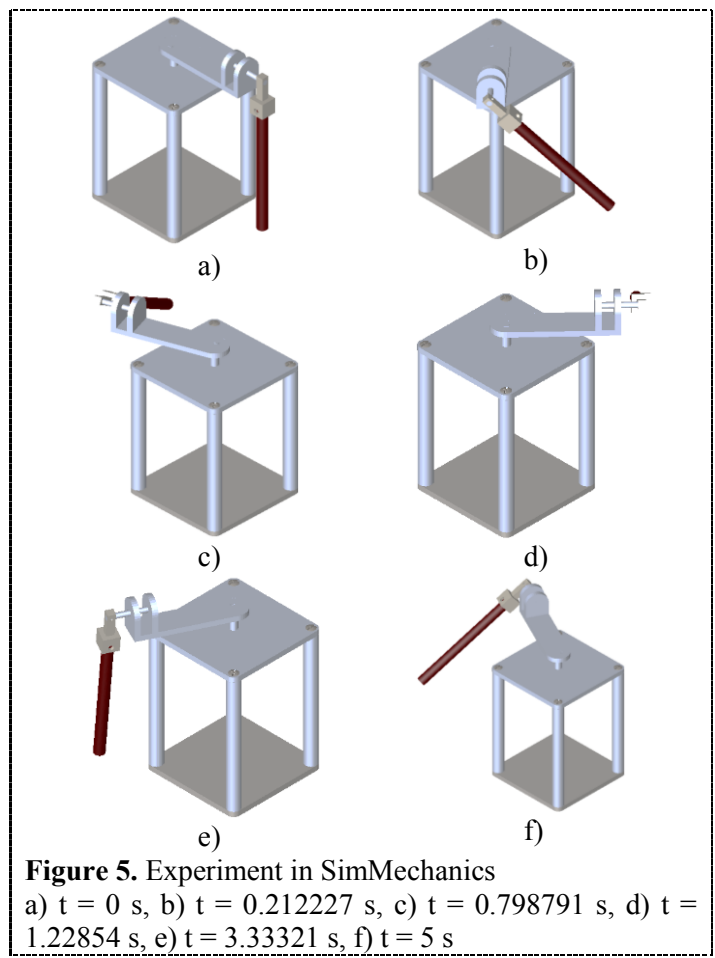

**Figure 5.** Experiment in SimMechanics
a) t = 0 s, b) t = 0.212227 s, c) t = 0.798791 s, d) t = 1.22854 s, e) t = 3.33321 s, f) t = 5 s

*6)* To export system solutions to Workspace, is need to add the blocks with the name To Workspace located in Simulink library, which will create a cell for each solution, with the respective data and time in which such data is obtained.

*7)* The implementation of the block diagram was performed to numerically solve the equations of the system, thus obtaining values $\theta_0, \theta_1, \dot{\theta}_0, \dot{\theta}_1$ along the defined time interval. The diagram is implemented as follows:

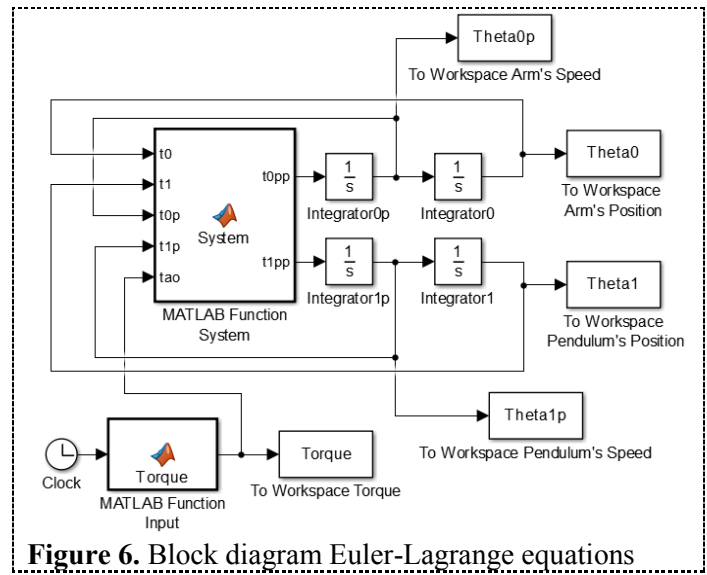

**Figure 6.** Block diagram Euler-Lagrange equations

Wherein the first block has the input applied to the system, which in this case is a torque step 0.2 second, which has the value 0.5 Nm which is the maximum torque provided by the engine, the block is called System contains the model equations in the state space, with 5 inputs ($\theta_0, \theta_1, \dot{\theta}_0, \dot{\theta}_1, \tau$) and 2 outputs ($\ddot{\theta}_0, \ddot{\theta}_1$). Internally, the block uses the ODE23tb method for solving equations.

Numerical data necessary to solve the equations of motion Euler-Lagrange are as follows:

**Table 2.** Numerical Variable System

| Physical characteristic | Symbology |
|---|---|
| Pendulum mass | $0.2866\ kg$ |
| Arm length | $0.201\ m$ |
| Pendulum length | $0.30997\ m$ |
| Location of the center of mass of the pendulum | $0.154985\ m$ |
| Gravity | $9.81\ m/s^2$ |
| Moment of inertia of the arm | $0.0052\ kg.m^2$ |
| Moment of inertia of the pendulum | $0.0023\ kg.m^2$ |

Was made an approximation in the calculation of the moments of inertia, both links are taken as constant circular bars and invariant mass,

$$I_0 = \frac{1}{3} m_0 L_0{}^2 \tag{36}$$

$$I_1 = \frac{1}{12} m_1 L_1{}^2 \tag{37}$$

Where equation (36) is the moment of inertia of the arm measured from the end connected to the motor shaft to the opposite end. While equation (37) is the pendulum moment of inertia measured pendulum from the center of mass.

*8)* The way to export the data to Workspace of block diagram above is performed with the same aggregate block in step *6)*.

*9)* Taking the exported data in points *6)* and *8)* the following graphs were made.

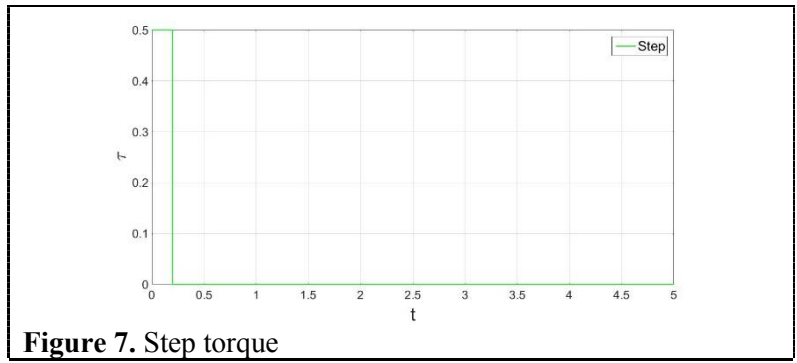

**Figure 7.** Step torque

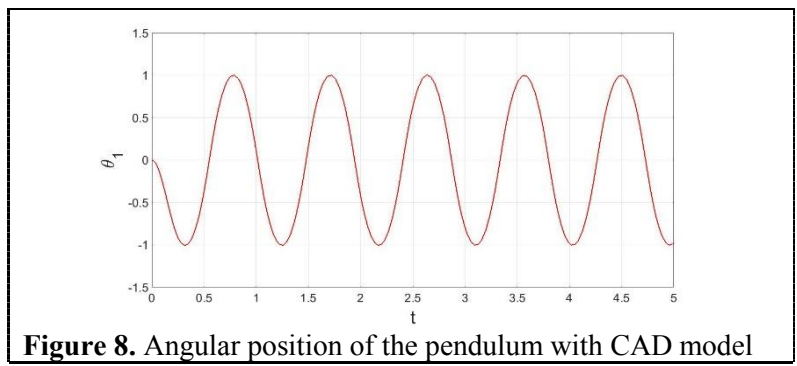

**Figure 8.** Angular position of the pendulum with CAD model

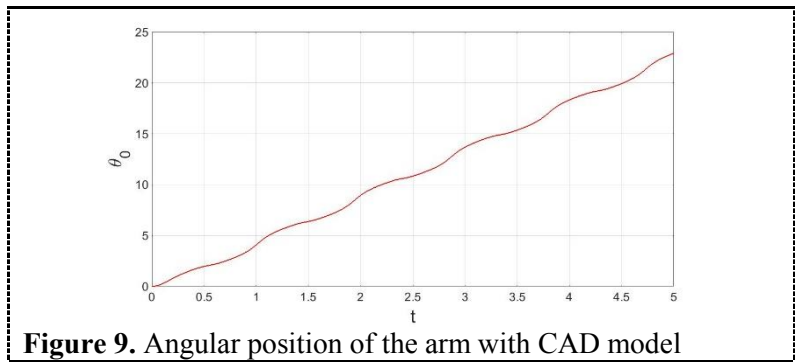

**Figure 9.** Angular position of the arm with CAD model

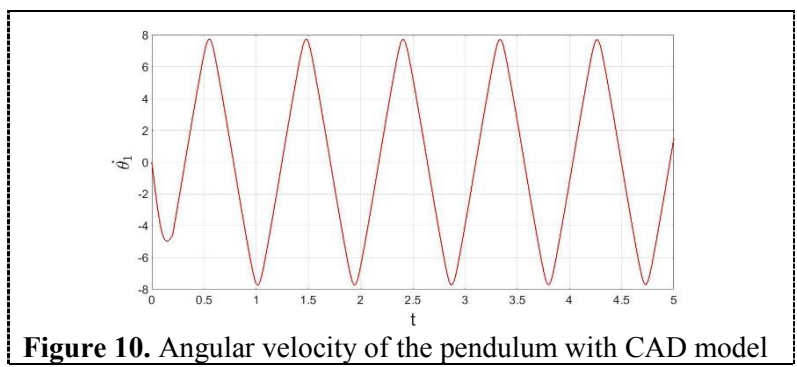

**Figure 10.** Angular velocity of the pendulum with CAD model

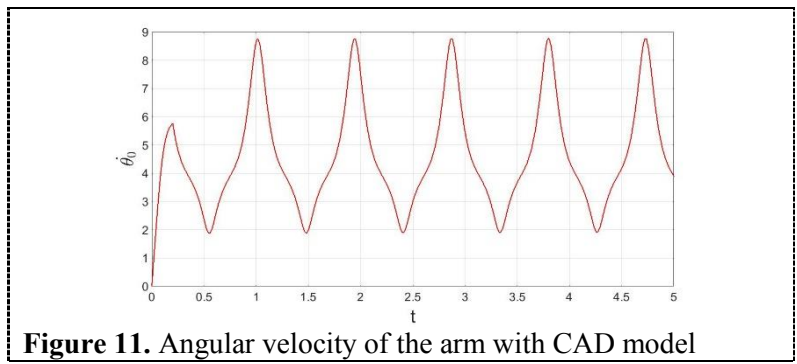

**Figure 11.** Angular velocity of the arm with CAD model

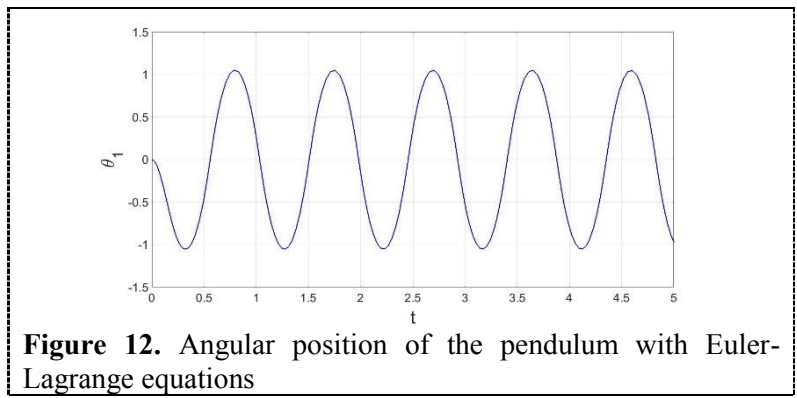

**Figure 12.** Angular position of the pendulum with Euler-Lagrange equations

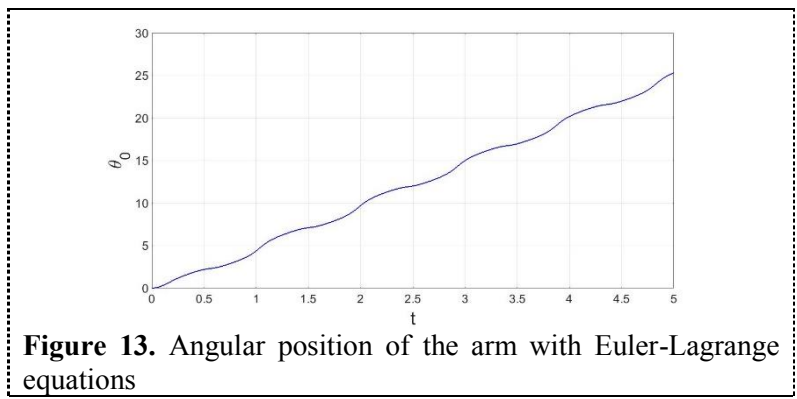

**Figure 13.** Angular position of the arm with Euler-Lagrange equations

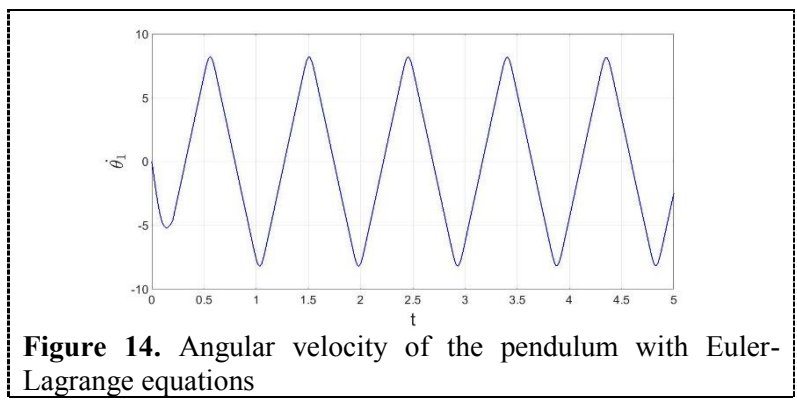

**Figure 14.** Angular velocity of the pendulum with Euler-Lagrange equations

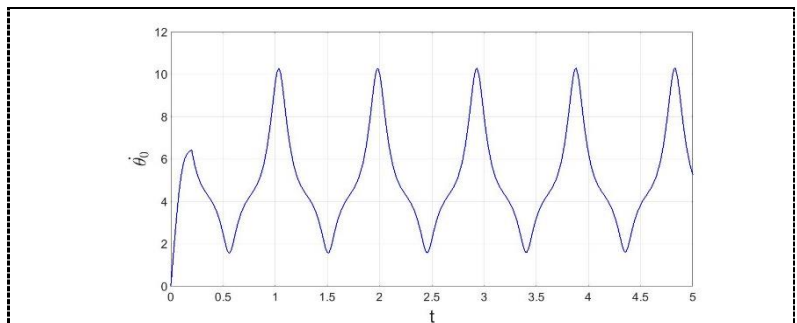

**Figure 15.** Angular velocity of the arm with Euler-Lagrange equations

It is noted that the system behavior is the same in both simulations. A superposition of results was performed by way of display the error between the CAD model simulated and the Euler-Lagrange equation, with the following results.

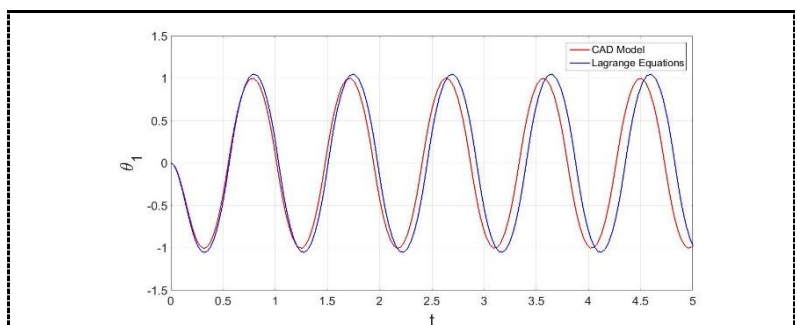

**Figure 16.** Comparison results angular position of the pendulum

Figure 16 allows viewing behavior overlapping displacement of the center of mass in the pendulum. The red line represents the CAD model while the blue is the result of $\theta_1$ graph the solution over time.

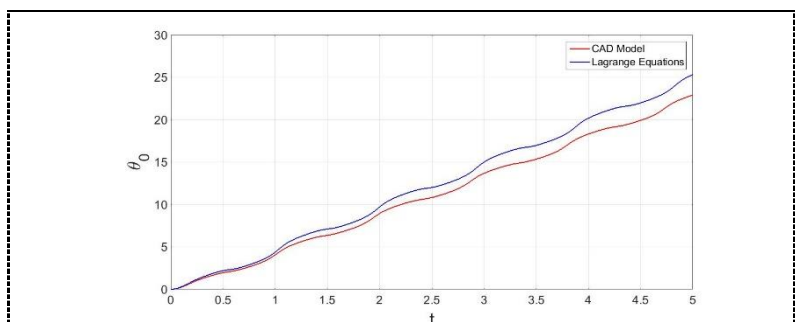

**Figure 17.** Comparison results angular position of the arm

Figure 17 is overlaying the results to simulate and plot $\theta_0$ function of time, where the red line is the result of simulating in SimMechanics the CAD model and the blue line is the solution of the Euler-Lagrange equation.

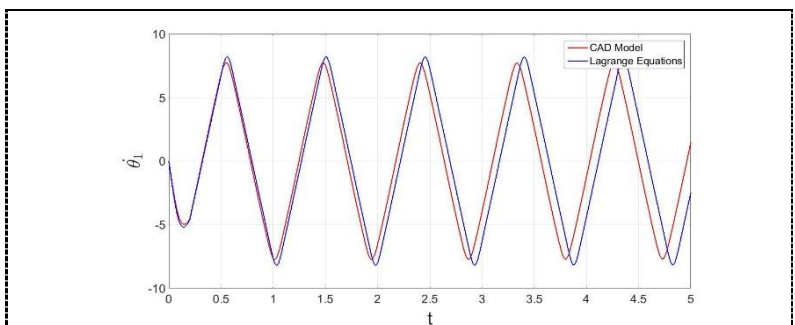

**Figure 18.** Comparison results angular velocity of the pendulum

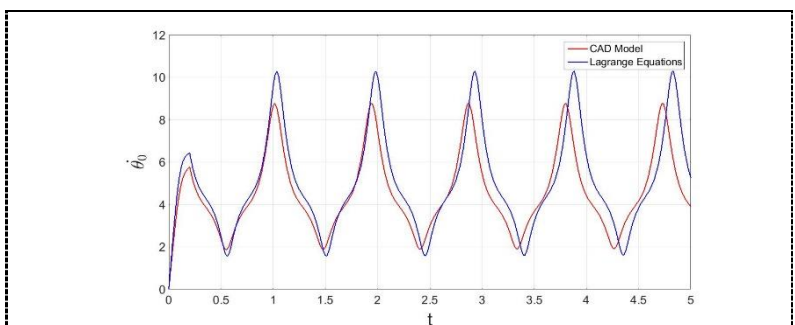

**Figure 19.** Comparison results angular velocity of the arm

Similarly as with Figure 16 and 17 the Figure 18 and 19 illustrate the behavior of the angular velocity with respect to time for the pendulum and the arm respectively.

It is important to note that the graphs do not have a zero error due to the approximations that were made to the calculation of the moments of inertia of each link, the difference is that the SolidWorks software performs a more accurate calculation of the moments of inertia, because the software considers the geometry of each link.

Then, the pendulum trajectory analysis shown, which part of the potential energy in the center of mass as shown in equation (9),

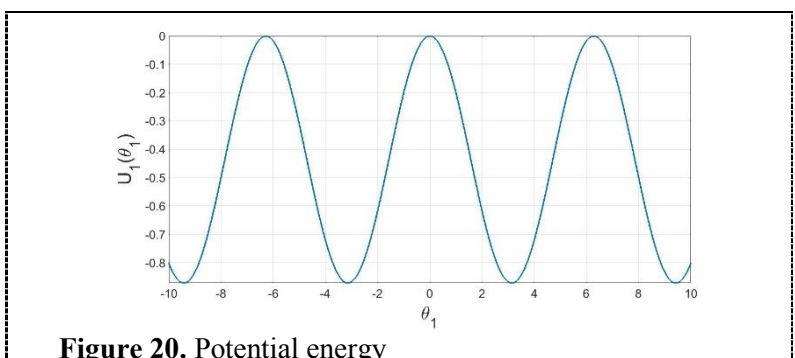

**Figure 20.** Potential energy

When all paths of the pendulum are collected the phase diagram, which also shows which are the critical points of the system, in this case we know that the pendulum has two critical points, a stable point at which the pendulum is in its obtained hanging position and unstable point in this case refers to a metastable point is satisfied when the pendulum is reversed. Graphically you can get the critical points in the system, we can first obtain which is the equation for calculating the minimum by Figure 20, it can be seen that the minimum occurs when the wave goes from a negative slope to a positive slope, therefore, we are obtained minimum $\pm\pi i$ where $i$ must be odd. Similarly one can find the maximum, these occur when moving from a positive slope to a negative, therefore the maximum are present in $\pm\pi i$ where $i$ must be equal to 0 or couple.

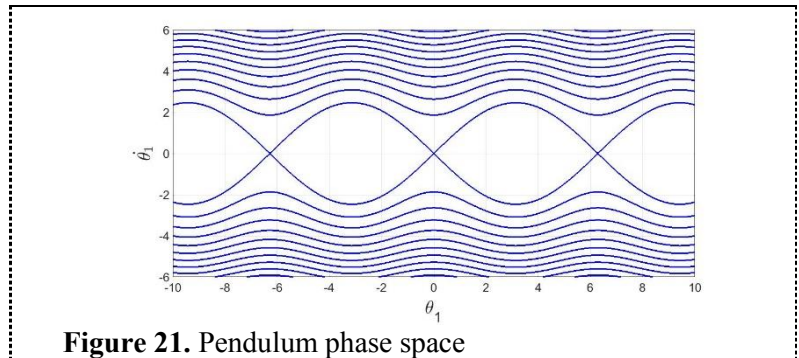

**Figure 21.** Pendulum phase space

Physically each point on the graph above represents a possible system status, explicitly state in which the pendulum would.

As mentioned above the pendulum has two equilibrium points, however, in the phase space endless these points are. This is because there is shown the position like a straight or $R$ space when the real space is a circle that is not topologically equals $R$. Therefore, if space representing geometric positioning system is not topologically equivalent to $R$, for subsequent phase space will not be in $R^2$, since this is the Cartesian product between the geometric space and space representing the speed is generally $R$. Topologically speaking the pendulum has a geometric space $S1$ which is equivalent to a circle and a space velocity $R$, therefore, the phase space $S1 \times R$ equivalent to a cylinder and not to $R^2$. The correct graphic process should be done on a cylinder where its circumference is the geometric representation of the pendulum, this is equivalent to the set of possible positions of the center of mass, while the position along its axis represents the velocity of the center mass, this cylinder is topologically correct space phase. Touring the circumference of the cylinder we can see that there is only one minimum and one maximum diametrically opposed. In Figure 20 you could take energy levels were the result of drawing lines parallel to the axis of the $\theta_1$, straight position, well, now those levels may be taken with circles surrounding the cylinder.

## 8. Conclusions

The Euler-Lagrange formalism allows dynamic modeling of rotational inverted pendulum in a simple way, thanks to the classical mechanics allows us to work on a climb.

This paper has presented step by step dynamic modeling of the proposed system, in addition to the respective simulation of the solution of the equations and further validation of the results using a simulation CAD model designed in SolidWorks exported in SimMechanics Matlab.

This work differs from others through the use of SimMechanics extension for simulation CAD model, this tool is easy to use and reliable enough, this is an advantage for simulations of 3D models since it is not necessary to create platforms for simulation. It also allows working in different ways on the same software, which in this case was Matlab, which allowed us to work with the CAD model and the equations of motion in the same environment (Simulink).

Something extra that is presented in this article is the analysis as to the trajectories of the pendulum, the phase space is topologically correct shown, this being a good contribution to future dynamic, kinematic and physical studies of the system.

**References**
[1]  Gómez F 2002 *"Control de sistemas no lineales basado en la estructura hamiltoniana"* tesis doctoral, Universidad de Sevilla España
[2]  Acosta J 2010 *"Furuta's pendulum: A conservative nonlinear model for theory and practice"* Mathematical Problems in Engineering
[3]  Cruz M,Silva R, Merlo C, Villarreal M, Muñoz D and Hernandez V 2014 *"Modeling and Construction of a Furuta Pendulum Prototype"* International Conference on Mechatronics, Electronics and Automotive Engineering
[4]  Blajer W and Kolodziejczyk K 2008 *Modeling of underactuated mechanical systems in partly specified motion* Journal of Theoretical and applied mechanics
[5]  Kajita S, Morisawa M, Miura K, Nakaoka S, Harada K, Kaneko K, et al 2010 *"Biped walking stabilization based on linear inverted pendulum tracking"* In: Proceedings of the IEEE/RSJ 2010 International Conference on Intelligent Robots and Systems, Taipei, pp. 4489-449
[6]  Younis W and Abdelati M. 2009. Design and implementation of an experimental segway model, AIP Conference Proceedings, 1107, p. 350-354
[7]  VIguria A, Cano, Fiacchini M, Prieto A, Vela B, Rubio F, Aracil J, and  Canudas de Wit C *"PPCAR (Personal Pendulum Car): Vehículo Basado en Péndulo Invertido"*. Departament Ingeniería de Sistemas y Automática, Universidad de Sevilla, España. Laboratoire d'Automatique de Grenoble (CNRS-LAG), Francia.
[8]  Quanser Innovate Educate, "The Rotatory Control Lab a Modular Single Source Solution You Can Control", www.quanser.com
[9]  Toro R 2009 *"Diseño y Control de un Péndulo Furuta para su Utilización en las Aulas de Clase de la Universidad EAFIT"*, Proyecto de grado para su utilización en las aulas de clase de la Universidad EAFIT. Universidad EAFIT, escuelas de ingenierías, departamento de ingeniería mecánica. Medellin, pp. 88
[10] Huang Ch, Cheng F and Ju Y 2010 *"Robust Control of a Furuta Pendulum"* SICE Annual Conference 2010, The Grand Hotel, Taipei, Taiwan
[11] Fedák V, Ďurovský F and Üveges F 2014 *"Analysis of Robotic System Motion in SimMechanics and MATLAB GUI Environment"*, Department of Electrical Engineering and Mechatronics, FEEaI, Technical University of Kosice, Slovakia
[12] Marion J 1998 *"Dinámica clásica de las partículas y sistemas"*. University of Maryland College Park. Academic Press, New York y London. Editorial Reverté, S. A, pp 643
[13] Conejo C *"Control Robusto H de Sensibilidad Mezclada Aplicado a Sistemas Lineales Invariantes en el Tiempo Subactuados"* tesis al grado de maestro en ciencias de la ingeniería eléctrica con orientación en control automático, Universidad Autónoma de Nuevo León, Facultad de Ingeniería Mecánica y Eléctrica. División de Estudios de Posgrado pp. 112
[14] Kelly R and Santibáñez V 2003 *"Control de Movimiento de Robots Manipuladores"*, PEARSON EDUCATION, S.A., Madrid, pp. 59-113